\documentclass{article}
\def\bsa#1#2\esa{\begin{subequations}\label{#1}
		\begin{align}#2\end{align} \end{subequations}}
\def\ba#1\ea{\begin{align}#1\end{align}}
\usepackage{graphicx,geometry}
\usepackage{epstopdf, epsfig,natbib,authblk}

\usepackage{amsmath,amssymb,graphicx,xcolor,mathptmx} 
\def\lap{\nabla^2}\def\g{\nabla}\def\f{\frac}\def\b{\mathbf}\def\p{\partial}
\def\e{\textrm{e}}

\def\D{\mathcal{D}}
\def\E{\mathcal{E}}
\def\F{\mathcal{F}}
\def\O{\mathcal{O}}
\def\u{\textbf{u}}
\def\x{\textbf{x}}
\def\om{\omega}
\def\A{\mathcal{A}}
\def\B{\mathcal{B}}
\def\G{\mathcal{G}}

\usepackage{amsmath}
\usepackage{color}
\usepackage{upgreek}

\definecolor{darkblue}{RGB}{83,0,93}

\usepackage{float}

\def\nn{\nonumber}
\def\d{\textrm d}

\def\ep{\epsilon}
\def\ba#1\ea{\begin{align}#1\end{align}}
\def\!#1\!{}

\def\bsa#1#2\esa{\begin{subequations}\label{#1}
\begin{align}#2\end{align} \end{subequations}}

\def\lp{\left(}
\def\rp{\right)}
\def\lb{\left[}
\def\rb{\right]}
\def\lcb{\left\{}
\def\rcb{\right\}}

\def\O{{\mathcal{O}}}

\title{Inherently Unstable Internal Gravity Waves due to Resonant Harmonic Generation}

\author{Y. Liang, Ahmad Zareei, M.-Reza Alam\thanks{reza.alam@berkeley.edu}}

\affil{\textit{Department of Mechanical Engineering, University of California Berkeley, Berkeley, CA 94720, USA}}
\date{}
\begin{document}
\maketitle

\begin{abstract}
Here we show that there exist internal gravity waves that are inherently unstable, that is, they cannot exist in nature for a long time. The instability mechanism is a one-way (irreversible) harmonic-generation resonance that permanently transfers the energy of an internal wave to its higher harmonics. We show that, in fact, there are countably infinite number of such unstable waves. For the harmonic-generation resonance to take place, nonlinear terms in the free surface boundary condition play a pivotal role, and the instability does not obtain for a linearly-stratified fluid if a simplified boundary condition such as rigid lid or linear form is employed. Harmonic-generation resonance presented here also provides a mechanism for the transfer of the energy of the internal waves to the higher-frequency part of the spectrum where internal waves are more prone to breaking, hence losing energy to turbulence and heat and contributing to oceanic mixing. 
\end{abstract}



\section{Introduction}

Internal gravity waves, outcome of perpetually agitated density-stratified oceans, are known to play a critical role in the dynamics of our planet's energy balance: they absorb energy to  form, carry energy over long distances as they propagate, and release energy where they break \cite[][]{Staquet2002}. The latter phenomenon usually gives rise to considerable mixing \cite[c.f.][]{Ferrari2008} whereby nutrients also get distributed, which is vital for a wide range of marine life \cite[][]{Boyd2007,Harris2012}. 

More than a century long research has shed a lot of light on various features of internal gravity waves. Nevertheless, many aspects of their inception and fate is yet not well understood \cite[e.g.][]{Alford2015}. Specifically, the precise mechanism that transfers energy from longer waves to the high-frequency part of the spectrum, where internal waves are more prone to breaking, is yet a matter of dispute. Aside from linear processes such as interaction of internal waves with the seabed topography and sloped continental shelves \cite[e.g.][]{Zhang2008\!,Zhang2014\!}, several nonlinear instability mechanisms have also been put forward. For instance, we now know that internal waves may undergo instability due to triad resonance \cite[][]{Davis1967,Hasselmann1967,Thorpe1968,Mccomas1977,Jiang2009,Scolan2013,Alam2009a\!,Alam2009b\!,Alam2011b\!,Alam2012b\!,Wunsch2015}. 
%
 All discovered destabilizing mechanisms for an internal wave (few named above), however, have one thing in common that they require some type of perturbations in order to get initiated. These perturbations can come from, for instance, seabed corrugations or presence of other waves forming resonance triads.


Here, we show that there are internal gravity waves in the ocean that are \textit{inherently} unstable, that is, they simply cannot sustain their form. 
%
Through the mechanism studied here, specific internal waves \textit{naturally} (without requiring any perturbation) give up their energy \textit{permanently} to their higher harmonics through a one-way irreversible harmonic-generation resonance mechanism.

\section{Governing Equations and the Dispersion Relation}
Consider the propagation of internal waves in an inviscid, incompressible, adiabatic and stably stratified fluid of density $\rho(x,y,z,t)$, bounded by a free surface on the top and a rigid seafloor at the depth $h$. Let's define a Cartesian coordinate system with $x,y$-axes on the mean free surface and $z$-axis positive upward. Newton's second law, conservation of mass, and conservation of energy provide five equations for the evolution of the components of the velocity vector $\b{u} = \{u, v, w\}$, density $\rho$, and the pressure $p$. These governing equations together with three boundary conditions (two kinematic boundary conditions on the free surface and the seabed, and one dynamic boundary condition on the free surface) uniquely determine the five unknowns and the surface elevation $\eta(x,y,t)$ \cite[e.g.][]{Thorpe1966}.

We assume internal waves are small perturbations to a stable background state at equilibrium. Therefore, density can be written as $\rho(x,y,z,t) = \bar{\rho}(z)+ \rho' (x, y, z,t)$ where $\bar{\rho}(z)$ is the background (unperturbed) density. Similarly, we define a pressure perturbation $p'$ via $p=\bar p(z)+p'(x,y,z,t)$ such that $d\bar p(z)/dz=-\bar{\rho}(z) g$. With some standard manipulation, the governing equations can be written in terms of either of the five variables involved in this problem. We choose to write the equation, as is customary, in terms of the vertical component of the velocity, $w$. These equations then read  \cite[see e.g.][or Appendix]{Thorpe1966}
\bsa{100}
&\f{\p^2}{\p t^2}\nabla^2w+N^2\nabla^2_Hw=\E(\u,\rho'),&-h<z<\eta,\label{101}\\
&\f{\p^3 w}{\p z \p t^2}-g\nabla^2_H w = \F(\u,p',\eta), & z=0,\label{102}\\
& w=0, &z=-h.\label{103}
\esa
where $\nabla^2_H=\p^2/\p x^2+\p^2/\p y^2$ is the horizontal Laplacian, $N^2=-{g/\rho_0~ \d \bar{\rho}(z)}/{ \d z}$ is the Brunt-V{\"a}is{\"a}l{\"a} frequency in which $\rho_0=\bar{\rho}(z=0)$ is the density on the free surface,  and $\E,\F$ are nonlinear functions of their arguments.

To perform a perturbation analysis, we assume that the solution to \eqref{100} can be expressed in terms of a convergent series, i.e.
\begin{align}\label{110}
w(\x,t)=\ep w^{(1)}(\x,t)+\ep^2 w^{(2)}(\x,t)+\O(\ep^3),
\end{align}
where $\ep\ll1$ is a measure of steepness of the waves involved and $w^{(i)}\sim \O(1)$. Similar expressions hold for $u,v,\rho'$ and $p'$. Substituting \eqref{110} into \eqref{100} and collecting terms of the same magnitude, then at the leading order $\O(\ep)$ the linearized equations are obtained.

We focus our attention here on the two-dimensional problem with a linear mean density profile, i.e. $\bar{\rho}(z)=\rho_0 (1-a z)$  
 which gives a constant Brunt-V{\"a}is{\"a}l{\"a} frequency $N=\sqrt{ga}$  \cite[c.f. e.g.][]{Martin1972}. Looking for a progressive wave solution of the leading order (linearized) equation in the form $w^{(1)}=W(z)\sin({\bf k}\cdot {\bf x} -\omega t)$ the following dispersion relations result:

\begin{align}\label{2000}
  \hspace{-.70cm} \D(k,\omega)=
\begin{cases}
    \omega^2-\f{g k}{\sqrt{1-{N^2}/{\omega^2}}}\tanh\lp kh \sqrt{1-{N^2}/{\omega^2}}\rp=0,&  \omega>N\\
    \omega^2-\f{g k}{\sqrt{N^2/\omega^2-1}}\tan\lp kh \sqrt{N^2/\omega^2-1} \rp=0,              & \omega<N
\end{cases}
\end{align}

\begin{figure}
\centering
\includegraphics[width=8cm]{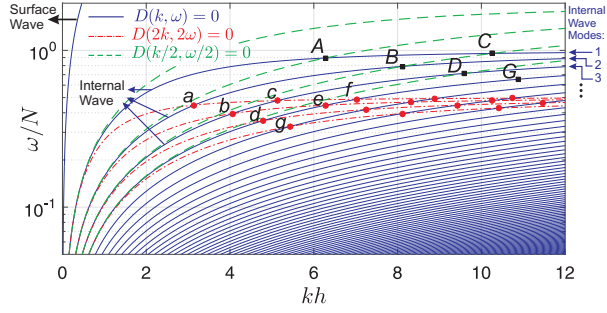}
\caption{Plot of the dimensionless frequency $\omega/N$ as a function of dimensionless wavenumber $kh$ of free internal waves (i.e. $\D(k,\omega)=0$) in a fluid of linearly stratified density $\rho(z) =\rho_0(1 - az)$, with $ah$ = 0.05. Associated with each wavenumber there is one surface wave and an infinite number of internal wave modes (blue solid-line branches). Frequency of internal waves cannot exceed the Brunt-V{\"a}is{\"a}l{\"a} frequency $N$, and all branches of the dispersion relation curve are capped at $\omega/N$=1. We also plot contours of $\D(2k,2\omega)$=0 (red dash-dotted lines) whose intersections with blue lines (shown by red circles) mark waves whose second harmonics are also solutions to the dispersion relation. These second harmonics are at the intersections of contours of $\D(k/2,\omega/2)$=0 (green dashed lines) and $\D(k,\omega)$=0 and are marked by black squares. The second harmonic of the wave at ``\textbf{\textit{a}}" (mode 2), is the wave ``\textbf{\textit{A}}" (mode 1) and so on. Note that second harmonic waves are at least one mode lower than the original waves.}
\label{fig1}
\end{figure}

Solutions to the above dispersion relation identify permissible frequency and wavenumber of free propagating waves. Contours of $\D(k,\omega)=0$ are shown in figure \ref{fig1} in which we plot the dimensionless frequency $\omega/N$ as a function of dimensionless wavenumber $kh$ (blue solid curves). For $\omega>N$ only one solution exists in the first quadrant (with its mirrors in the other quadrants). This solution corresponds to a wave whose associated fluid particle motion is maximum near the free surface and decreases as the depth increases. Therefore this is basically a classical \textit{surface} wave which is a little perturbed because of stratification. For $\omega<N$ there is an infinite number of solutions to \eqref{2000}. The first member of this set, is the continuation of the surface wave branch (the left-most branch in figure \ref{fig1}), but the rest identify waves with associated fluid particle activities that are minimum near the free surface and the seabed, but gain one (or more) maximum/maxima somewhere inside the fluid domain. Therefore these branches show \textit{internal} waves. The number of maxima in the amplitude of velocity along the vertical water line determines the mode number of the branch (the first three are marked on the right-side of figure \ref{fig1} with arrows). A cut-off frequency $\omega/N$=1 sets an upper frequency limit for internal waves. It is to be noted that the dispersion relation \eqref{110}, although has a different form, is in fact graphically very close to the one under rigid lid assumption $\omega=N/\sqrt{1+(n\pi/kh)^2}$. But clearly the inclusion of the effect of the free surface in the former has changed its form.

\section{Harmonic Generation}

With the linear solution to \eqref{100} and its properties at hand, we move to the second order equation by collecting $\O(\ep^2)$ terms obtained from the substitution of \eqref{110} into \eqref{100}. The second order equation has the exact same form as of the leading order equation on its left-hand side, but with nonlinear terms, arising from nonlinear functions $\E,\F$, on its right-hand side. These nonlinear terms are multiplication of the linear solution (and its derivatives) and, it turns out that, they constitute forcing terms with wavenumber and frequency $(2k,2\om)$. Now if $\D(2k,2\om)$=0, then it means that these forcing terms have a harmonic which is the same as the natural harmonic of the linear system. This is a resonance scenario through which a new second order solution may emerge and may grow large enough to the extent that it becomes comparable to the leading order solution (beyond which the naive expansion \eqref{110} is not valid anymore). We would like to note a subtle point here that at the second order the right-side of \eqref{101} is identically zero \cite[see e.g.][]{Tabaei2005}. Therefore a potential harmonic-generation resonance is only possible through the nonlinear terms in the free surface boundary condition \eqref{102}. If a rigid-lid assumption or a linearized form of the free surface boundary condition is employed (which is usually the case in the investigation of internal waves) the resonance harmonic generation will simply not obtain unless a non-uniform stratification or Boussinesq terms are considered.

To see whether it is possible to satisfy the resonance condition required for the second harmonic of an internal wave to exist as a free propagating wave, we also plot in figure \ref{fig1} the contours of $\D(2k,2\om)$=0 (red dash-dotted curves) for first four internal wave modes. Intersections of these contours with the contours of $\D(k,\om)$=0 (solid blue curves) identify waves whose second harmonics are also solution to the dispersion relation. Some of these intersections are marked in figure \ref{fig1} by red circles and are identified by lowercase characters. It is easy to find the corresponding second harmonic by multiplying a designated (red circle) frequency and wavenumber by a factor of two, or alternatively by finding intersections of contours of $\D(k/2,\om/2)$=0 (green dashed curves) and $\D(k,\omega)$ (denoted by black squares and uppercase characters). Specifically, ``\textit{\textbf{A}}" is the second harmonic of $``\textit{\textbf{a}}"$, $``\textit{\textbf{B}}"$ is the second harmonic of $``\textit{\textbf{b}}"$ and so on. It is to be noted that the second harmonic of an internal wave always belongs to a lower mode than the mode of the original wave. For example, second harmonic of wave ``\textit{\textbf{a}}" (mode 2) is the wave ``\textit{\textbf{A}}" (mode 1), and second harmonic of wave ``\textit{\textbf{c}}" (mode 3) is the wave ``\textit{\textbf{C}}" (mode 1).

\begin{figure}
\centering
\includegraphics[width=8cm]{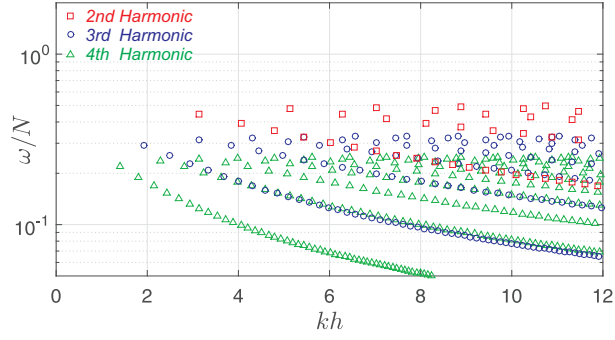}
\caption{There are countably infinite number of internal waves that are unstable to their second harmonic. Physically this means that specific incident internal waves of wavenumber and frequency ($k,\omega$) will give up their energy permanently (in a one-way irreversible process, c.f. equation \eqref{200}) to their second harmonic ($2k,2\om$). The necessary condition for this to happen is $\D(k,\om)=\D(2k,2\om)$=0. These waves, for the parameters of figure \ref{fig1}, are shown here by red squares. Similar story holds for another set of waves that are unstable to their third harmonic (blue circles, necessary codition $\D(k,\om)=\D(3k,3\om)$=0), and fourth harmonic (green triangles, $\D(k,\om)=\D(4k,4\om)$=0) and so on. Instability to higher harmonics are clearly much weaker when compared with the instability to the second harmonic.}
\label{fig2}
\end{figure}

A comprehensive collection of all internal waves (for $kh<$12) whose second harmonics are also free propagating waves is shown in figure \ref{fig2}. We are basically plotting in this figure all intersections points of contours of $\D(k,\om)=0$ and contours of $\D(2k,2\om)=0$. Clearly there will be an infinite (but countable) number of such solutions. Through the approach described above, a similar exercise can be performed for waves with their third harmonic being free waves. These solutions are marked by blue circles in figure \ref{fig2}, and waves with their fourth harmonics lie on the dispersion relation curves are shown by green triangles, and this search can continue indefinitely.

\section{Results and Discussions}

To determine the strength of the resonance (i.e. the rate of growth of the resonant wave), and the dynamics of the energy interplay between a wave and its second harmonic, here we perform a multiple scale perturbation analysis. The basic assumption is that the amplitude of the original wave and its second harmonic are both functions of spatial variables and time (i.e. $\x,t$), \textit{as well as} a slow spatial variable in the direction of propagation $x_1=\ep x$. Physically speaking, we allow the amplitude of both waves to \textit{slowly} vary as waves propagate. Mathematically speaking this is written as
\begin{align}\label{210}
w(x,x_1,z,t)=\ep w_1(x,x_1,z,t)+\ep^2 w_2(x,x_1,z,t)+\O(\ep^3)
\end{align}
where $w_i\sim\O(1)$. Expressions of the same form are assumed to hold for other variables $u,\rho',p'$ and $\eta$. Similar to regular perturbation methodology, described earlier in the paper to gain insight into the problem, by substituting \eqref{210} into the governing equation \eqref{100} and collecting terms of the same order in $\ep$. We assume the original wave with wavenumber and frequency $k,\omega$ has the amplitude $\A_1(x_1)$ and the amplitude of resonant second harmonic wave with wavenumber and frequency $2k,2\omega$ is $\B_2(x_1)$. At the second order, applying a compatibility condition (to avoid unbounded solutions) the following two equations governing spatial evolution of $\A_1(x_1)$ and $\B_2(x_1)$ emerge (see Appendix for details of derivation)
\bsa{200}
&\f{\d \B_2(x_1)}{\d x}=\alpha \A_1^2(x_1),\label{201}\\
&\f{\d \A_1(x_1)}{\d x}=\beta \A_1(x_1)\B_2(x_1),\label{202}
\esa
where
\begin{align}
&\alpha=-\f{6m_1^2\om\cos m_2h}{ g k(2m_2h+\sin 2 m_2 h) },\nn\\
&\beta=-\f{k[\sin m_2 h(4m_1\cos^2 m_1h -3m_1) -2m_2\cos m_2h\sin 2m_1h]}{2\omega(2m_1h+\sin 2m_1h)}. \nn
\end{align}
in which $m_1=k\sqrt{N^2-\omega^2}/\omega$ and $m_2=k\sqrt{N^2-4\omega^2}/\omega$.

\begin{figure}
\centering
\includegraphics[width=7cm]{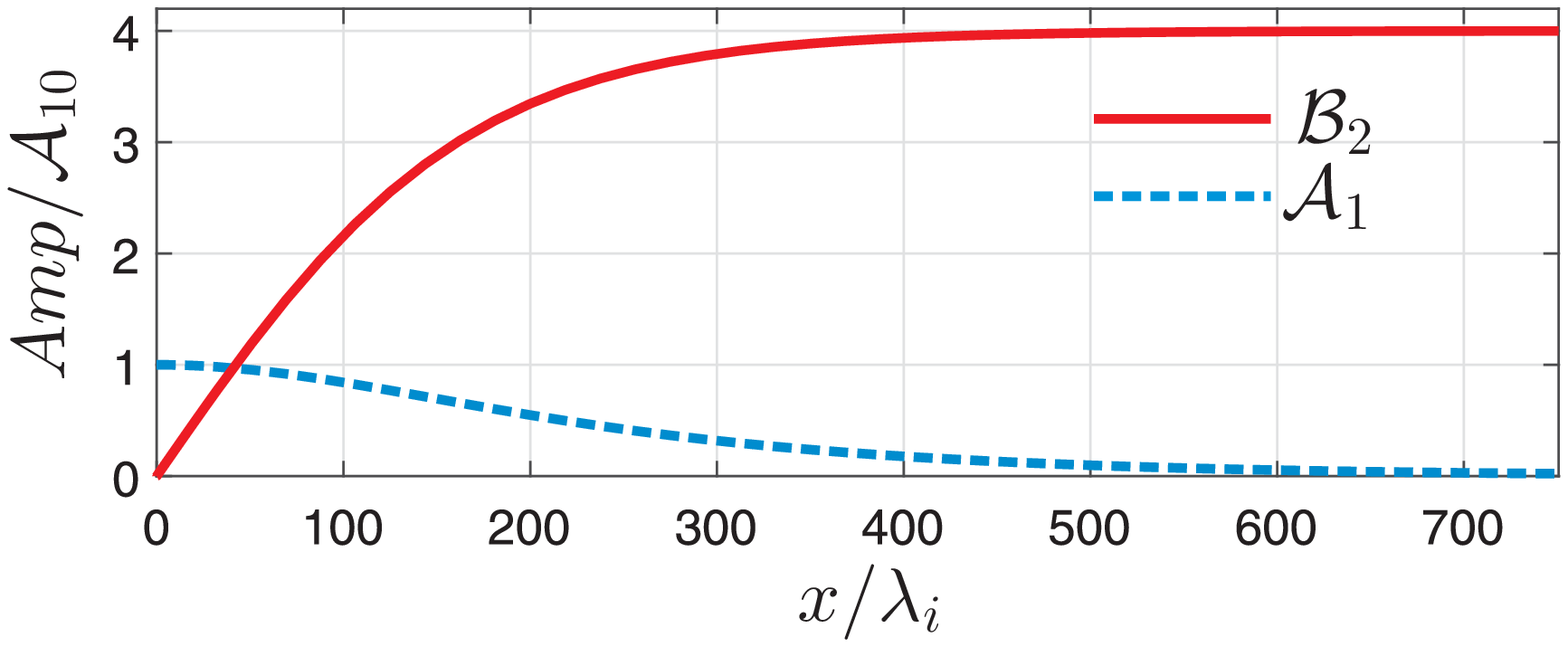}
\put(-205,82){(a)}\\
\includegraphics[width=7cm]{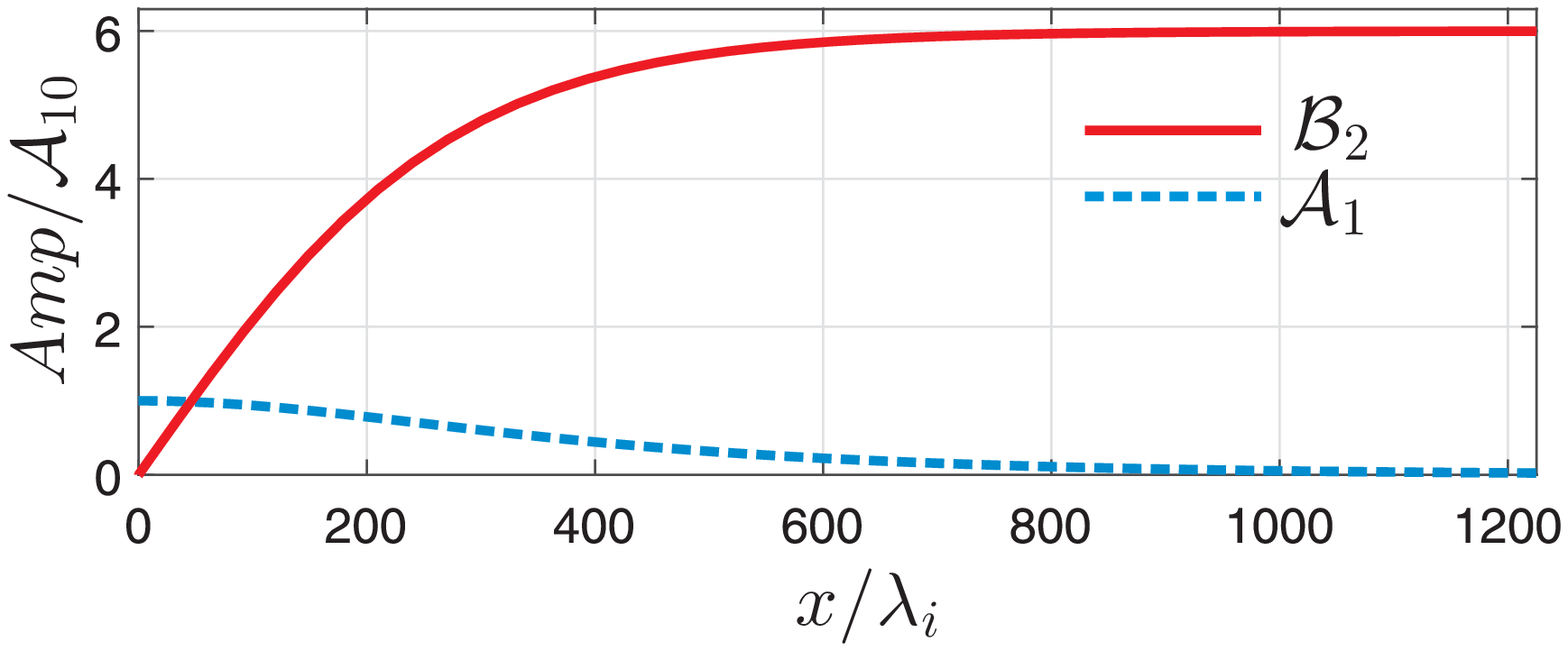}
\put(-205,82){(b)}\\
\includegraphics[width=7cm]{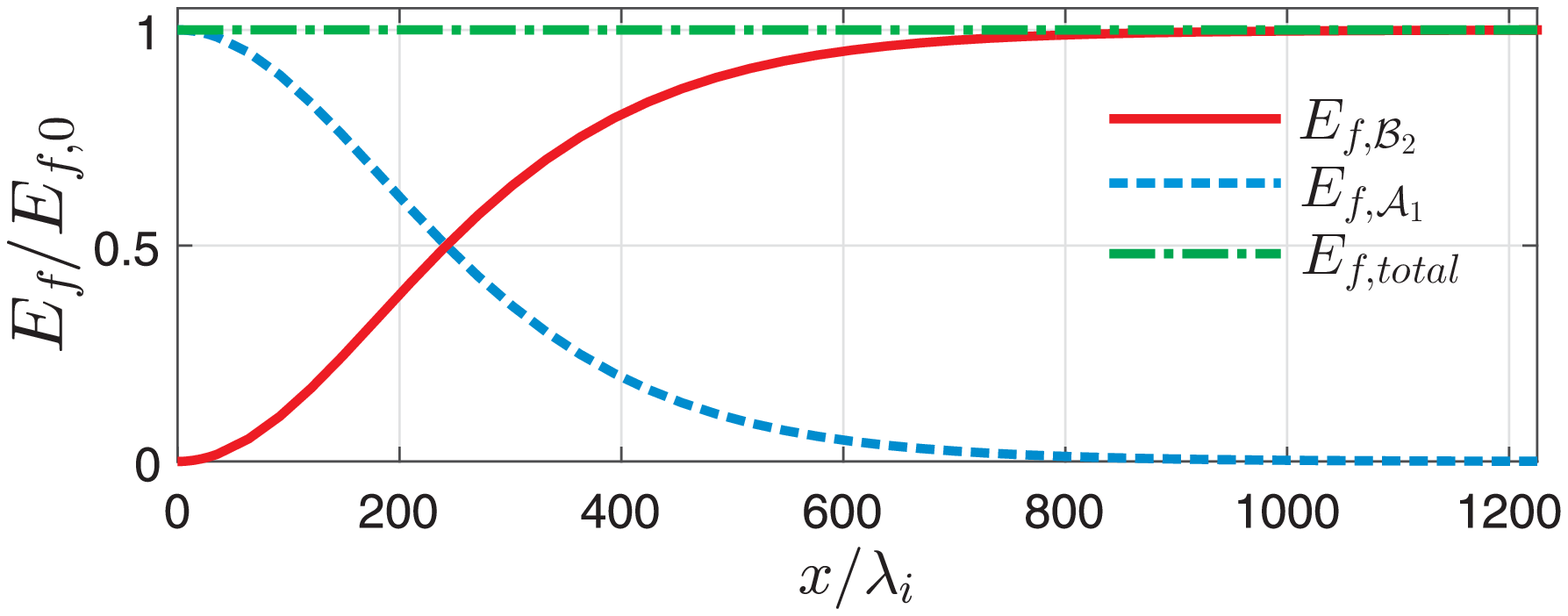}
\put(-200,73){(c)}
\caption{Spatial evolution of the amplitude of the original wave (blue dashed line) and its resonant second harmonic (red solid line) correspond to the point ``\textit{\textbf{a}}" in figure \ref{fig1} ($kh$=3.144772, $\omega/N$=0.4472136, $aA_{10}/N$=0.0023, fig. a), and the point ``\textit{\textbf{c}}" in figure \ref{fig1} ($kh$=5.132426, $\omega/N$=0.4780914, $aA_{10}/N$=0.0023, fig. b). In figure (a) energy goes from mode 2 to mode 1, whereas in figure (b) energy goes from mode 3 to mode 1. Figure (c) shows the energy flux of each wave as well as the sum of the energy fluxes (green dash-dotted line). As expected from energy conservation, the overall energy flux is unchanged in the domain of interaction. }
\label{fig3}
\end{figure}

Spatial evolution of the normalized amplitude of the original wave $\A_1/\A_{10}$ (where $\A_{10}=\A_1(x=0)$) and its second harmonic $\B_2/\A_{10}$ as a function of spatial distance of propagation $x/\lambda_i$ (where $\lambda_i=2\pi/k$= wavelength of the original wave)  is shown in figure \ref{fig3} respectively by blue-dashed line and solid-red line. In accordance with the case presented in figure \ref{fig1}, we choose $ah$=0.05 and if we consider that waves are propagating in a water of depth $h$=1 km, then $N$=0.02 rad/s. For this case, figure \ref{fig3}a corresponds to the point ``\textit{\textbf{a}}" ($kh$=3.144772,$\omega/N$=0.4472136) with its second harmonic at ``\textit{\textbf{A}}" in figure \ref{fig1}, and figure \ref{fig3}b corresponds to point ``\textit{\textbf{c}}" ($kh$=5.132426,$\omega/N$=0.4780914) with its second harmonic at ``\textit{\textbf{C}}". In the former case the interaction is (spatially) faster by a factor of about two, while in the latter case the relative amplitude of the resonant wave is higher. The most striking aspect of the solution is that the interaction is one-way. This can be seen from \eqref{201} in which, because $\A_1^2$ is always positive, then $\d \B_2/d x_1$ can never change sign. As a result the magnitude of $\B_2$ can only increase (the sign of $\alpha$ only contributes to 0 or $\pi$ radian phase shift to the wave), and that's why the direction of energy can never change. This is in contrast to typical triad resonance interactions \cite[e.g.][]{Alam2010,Alam2012c} and harmonic generation  in shallow water waves \cite[e.g.][]{Alam2007} where energy initially flows from original waves to resonant waves, but then when the amplitude of resonant waves is large enough the flow of energy reverses. Here, energy only goes from the original wave to the second harmonic and stays there permanently. The original wave will be gone forever.

Our multiple scales results conserve energy, as expected. In figure \ref{fig3}c (which corresponds to the case presented in figure \ref{fig3}b) we plot the energy flux ($E_f=E\times C_g$ where $E$ is the energy per unit area and $C_g$ is the group velocity) normalized by the energy flux of the original wave at the beginning ($E_{f,\A_{10}}$). Plotted are the normalized energy flux of the original wave $E_{f,\A_1}$ (blue-dashed line), the second harmonic $E_{f,\B_2}$, and the summation of the fluxes $E_{f,total}=E_{f,\A_1}+E_{f,\B_2}$. As expected from the conservation of energy the latter is constant and equal to unity.

To cross validate our results with direct simulation, we use the adaptive Navier-Stokes solver code SUNTANS (Stanford Unstructured Nonhydrostatic Terrain-following Adaptive Navier-Stokes Simulator \cite[][]{Fringer2006}). As a nonhydrostatic parallel ocean model with the capability to implement nonlinear free surface, SUNTANS has been validated and widely used in studying internal waves (and associated turbulence and mixing) in stratified waters \cite[][]{Zhang2011,Kang2010,Wang2011,Kang2012a,Walter2012,Zhang2011,Wang2011}. 

\begin{figure}
\centering
\includegraphics[width=6.5cm]{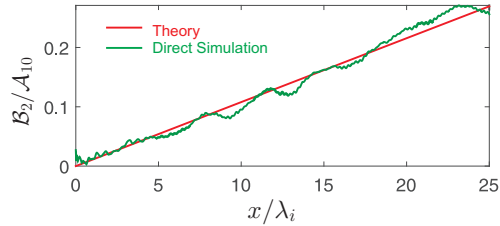}
\caption{Comparison of analytical results \eqref{200} with the direct simulation. The physical parameters used are: density gradient $a$=${1\e-3}$ $m^{-1}$, water depth h=100m, dimensionless wave number $kh$=3.147946, dimensionless frequency $\omega/N$=0.4472136, the amplitude of vertical velocity of the incident wave  $A_{10}$=0.1 m/s.
}
\label{fig4}
\end{figure}

Here we consider propagation of waves in a stratified water of density gradient $a=1\times10^{-3}$ m$^{-1}$, depth 100m and in a domain of horizontal extent 20km.  We choose $5\times10^3$ grid points in the $x$ direction and 100 layers in the $z$ direction. We specify the velocity of fluid particle at the left boundary to match that of desired incoming wave. Therefore the left boundary  that serves as the incoming wave boundary (wave maker). We set the right-side vertical boundary of the domain as the no-penetration and slip-free, and therefore it will act as a rigid vertical wall. We consider the case of incident internal wave of mode 2 ($kh$=3.147946, $\omega/N$=0.4472136, vertical velocity amplitude $w$=0.1m/s) with its second harmonic being a mode 1 wave. The time step $\delta t/T$=1/35000 where $T$ is the period of the incident wave. We run the simulation until the incident wave arrives at the wall, at which point the amplitude of both waves (incident and its second harmonics) have reached a steady state in the domain of interest $0<x/\lambda_i<20$. As is seen from figure \ref{fig4}, the curve of $\B_2/\A_{10}$ vs $x/\lambda_1$ gives an initial slope of 0.0107 which is the same as our theoretical prediction \eqref{200}. 

\begin{figure}
\centering
\includegraphics[width=8cm]{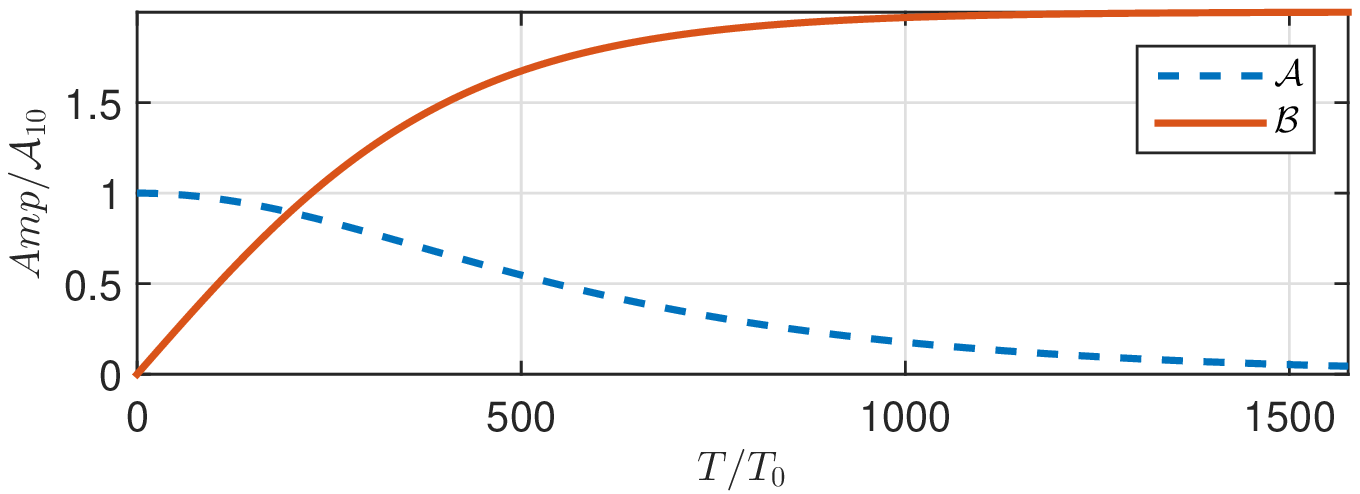}
\put(-245,82){(a)}\\
\includegraphics[width=8cm]{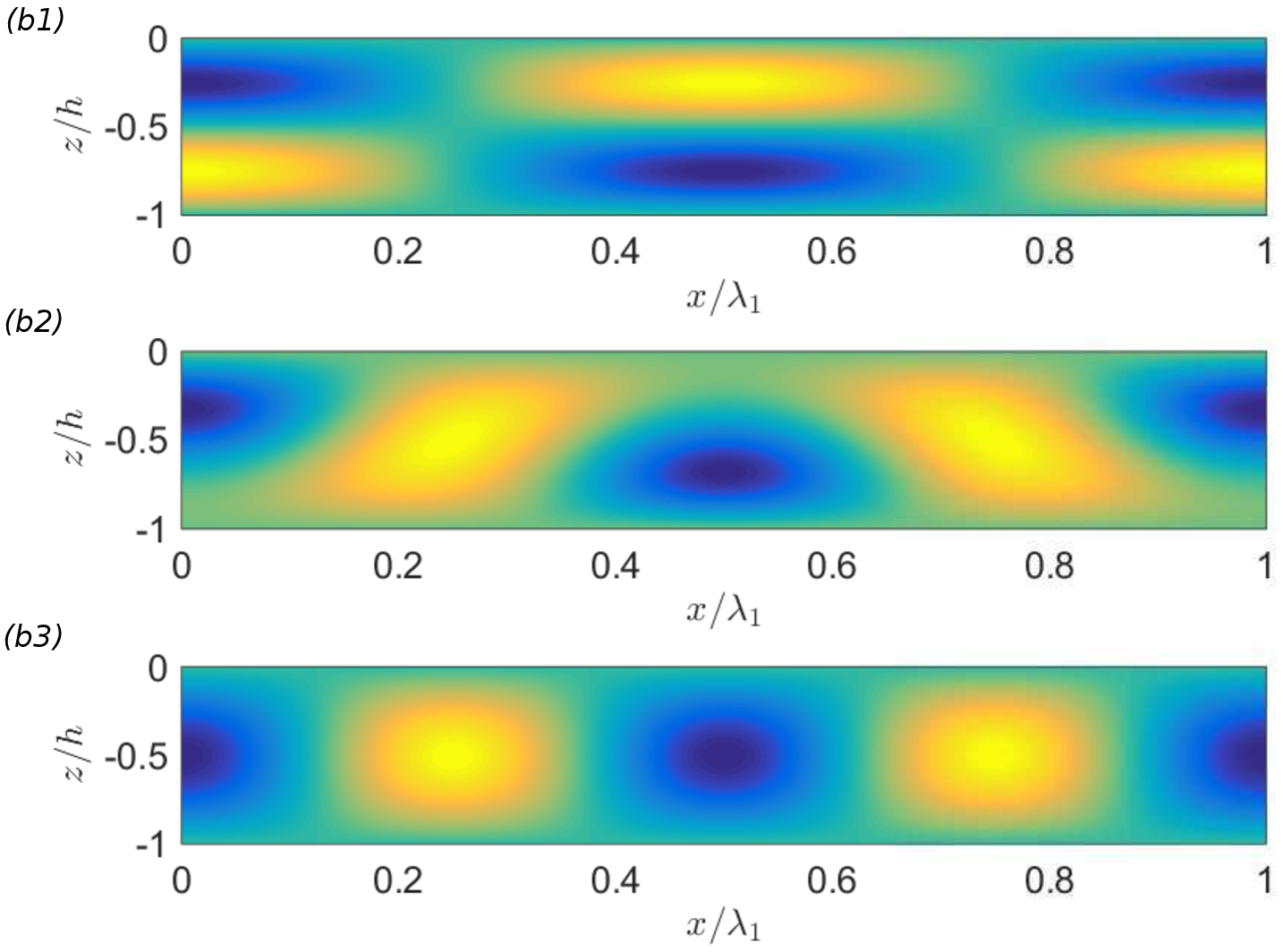}
\put(-245,82){(b)}\\
\caption{Time evolution of internal gravity waves that satisfy the conditions of harmonic generation. The physical parameters used here correspond to figure \ref{fig3}a. (a) Time evolution of the amplitudes of the two waves. $T_0$ is the period of the parent wave. (b) Snapshots of the wave field at $t=0$ (b1), $t=500T_0$ (b2), and $1500T_0$(b3). The presented mechanism initially results in an modulation of the primary wave, and eventually all the energy goes to the second harmonic.}
\label{fig42}
\end{figure}

To see how fast this instability evolves temporally, we present in figure \ref{fig42} results of the temporal evolution of the parent wave and its second harmonic, with the same physical parameters as in the case in figure  \ref{fig3}a.  The governing equation takes a similar form to \eqref{200} except the derivatives that are with respect to the time $t$ and clearly expressions for $\alpha$ and $\beta$ are different. The qualitative trend is as the spatial case, and the figure suggests that we need time $\sim\O(1000)$ times the period of the initial wave to see the majority of the energy transmitted to the second harmonic. With the chosen Brunt-V{\"a}is{\"a}l{\"a} frequency of $N=0.022$s$^{-1}$, the time scale is about 10 days, which is of the same order as that of parametric subharmonic instability of the M2 internal tide ($\sim$ 2-5 days) \cite[c.f. e.g.][]{Gerkema2006c,MacKinnon2005a}. This is somewhat expected as both mechanisms are \textit{second-order} resonance instabilities.

We would like to comment here that Parametric Subharmonic Instability is dependent upon perturbation waves in the domain in order to get started \cite[][]{Davis1967,muller1986}. In direct simulations, this is achieved by adding random noise to the simulation domain. In experimental studies the required noise already exists in the domain due to unavoidable imperfections. Hence, usually in experiments Parametric Subharmonic Instability is automatically obtained (similar to inevitable Benjamin-Feir instability) and destabilizes ``single" internal waves to a number of other subharmonic waves \cite[e.g.][]{Martin1972}. We would like to emphasize that the underlying mechanism of resonant harmonic generation studied here is different than that of Parametric Subharmonic Instability in that the former mechanism, among other characteristics, is an ``inherent" instability that does not require ambient perturbations to get started.

%




Three snapshots of the overall vertical velocity $w$ are shown in figure \ref{fig42}b that highlights the horizontal and vertical structures of the wave field at different times. Energy goes from the original internal wave (b1) slowly to its higher harmonics, first leading to modulation of the original wave (b2), but eventually the entire energy is at the second harmonic (b3) whose frequency and horizontal wavenumber are double the horizontal frequency and wavenumber of the initial wave. The interaction is only between the initial wave and its second harmonic, and no cascading to different waves is ensued.

\begin{figure}
\centering
\includegraphics[width=8cm]{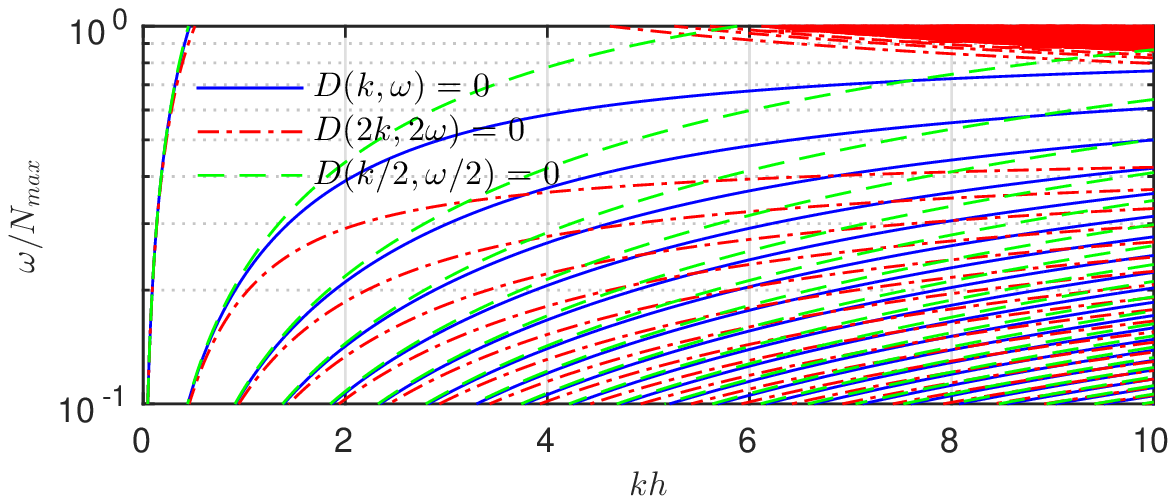}
\put(-235,82){(a)}\\
\vspace{0.4cm}
\includegraphics[width=8cm]{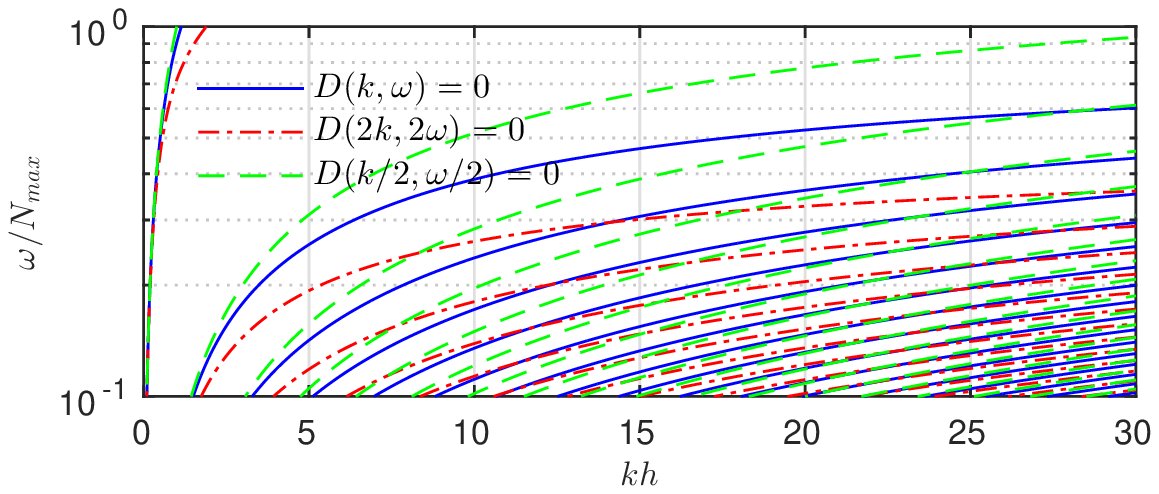}
\put(-235,82){(b)}\\
\caption{Plots of $\D(k,\omega)=0$, $\D(2k,2\omega)=0$ and $\D(k/2,\omega/2)=0$ for parabolic and exponential density profiles. (a) The density is $\rho(z)=\rho_0 (1+a z^2)$, where $a=1\times10^{-5} \text{m}^{-2}$. Water depth is $h=100$m. The Brunt-V{\"a}is{\"a}l{\"a} frequency $N=\sqrt{-2gaz}$. $N_\text{max}=\sqrt{2gah}$. (b) The density is $\rho(z)=\rho_0 [1+\delta-\delta \exp(az)]$, where $a=0.1 \text{m}^{-1}, \delta=0.1$. Water depth is $h=100$m. The Brunt-V{\"a}is{\"a}l{\"a} frequency $N=\sqrt{ga\delta\exp(az)}$. $N_\text{max}=\sqrt{ga\delta}$.}
\label{fig5}
\end{figure}

We would like to note that the harmonic generation mechanism proposed here occurs for non-uniformly stratified ocean as well. As pointed out previously, a necessary condition for this to occur is that for a given wave of frequency $\omega$ and wavenumber $k$, the two conditions $\D(k,\omega)=0$ and $\D(2k,2\omega)=0$ are satisfied. We present two examples in figure \ref{fig5}  for two different density profiles: parabolic $\rho(z)=\rho_0 (1+a z^2)$, where $a=1\times10^{-5} \text{m}^{-2}$ (figure \ref{fig5}a), and exponential $\rho(z)=\rho_0 [1+\delta-\delta \exp(az)]$, where $a=0.1 \text{m}^{-1}, \delta=0.1$ (figure \ref{fig5}b).  The Brunt-V{\"a}is{\"a}l{\"a} frequencies increase with depth in one case and decrease in the other. It can been seen from figure \ref{fig5} that there are, as well, infinitely countable intersections (i.e., solutions) between the two sets of curves. In these cases, contrary to the constant $N$ case, the linear wave solution does not satisfy the nonlinear Boussinesq-Euler equation and hence inclusion of the nonlinear free surface condition is not necessary for the harmonic generation to appear. The analysis for these two cases can be carried out by following the same logic presented in the appendix, although mathematically more involved and closed-form explicit solutions will be tedious, if not impossible, to obtain. 

As discussed above, when density profile is not uniform and hence $N$ is not constant, two sources contribute to the harmonic generation: 1- nonlinearities in the free surface boundary condition, i.e. right-hand side of equation \eqref{3002}, and  2- nonlinear terms in the momentum equation, i.e. right-hand side of equation \eqref{3001}. As presented before, for a constant $N$ case the latter source is absent. For nonuniform stratifications, it is of interest to evaluate the relative importance of these two contributors. The motivation is to see whether the classical rigid-lid assumption would be a good approximation in estimating energy exchange due to harmonic generation when the density profile is non-uniform (we already presented this is not the case for a uniform stratification). 

We comment on this question briefly by considering the exponential density profile as it is representative of actual pycnoclines. For such a density profile, the vertical structure of an internal wave $W(z)$ are obtained as
\ba
W(z)=-{\frac {{{\mathit J}_{\nu}\left(\mu\right)}{{\mathit Y}_{\nu}\left(p\right)}-
{{\mathit Y}_{\nu}\left(\mu\right)}{{\mathit J}_{\nu}\left(p\right)}}{{{\mathit Y}
_{\nu}\left(\mu\right)}}}
\ea
where, ${\mathit J}$ and ${\mathit Y}$ are Bessel function of the first and second kinds respectively, and 
\ba
\nu=\f{2k}{a}, ~~~\mu=\frac {2k\sqrt {g\delta}{{\rm e}^{-1/2\,ah}}}{\omega\,
\sqrt {a}},~~~p=\frac {2k\sqrt {g\delta}{{\rm e}^{1/2\,az}}}{\omega\,\sqrt 
{a}}.
\ea
By substituting this equation into the right-hand side of equations \eqref{3001} and \eqref{3002} the relative importance of the right-hand side of the two equations are obtained. 

 We consider a wave with the wave number and frequency $kh=14.14$ and $\omega/N_\text{max}=0.30$, taken from one solution in figure \ref{fig5}(b). It turns out that, in this example and for instance (at $z=-20$m), the relative magnitude of nonlinear terms on the right-hand side of the Boussinesq-Euler equation to those of free surface boundary condition (i.e. \eqref{3001} to \eqref{3002}) is about 30\%. This means that the contribution of free surface boundary condition to the harmonic generation may be multiple times larger than the nonlinear terms in the momentum equation. We would like to emphasize that the resonant harmonic generation studied here is different from \textit{non-resonant} harmonic generation that obtains in non-uniform stratifications \cite[e.g.][]{Sutherland2016}. The former occurs for specific internal waves, transfers the entire energy to the second harmonic and is a one-way process, whereas the latter obtains for all internal waves but since it is non-resonant only transfers a portion of energy from the initial wave to the superharmonics.

Similarly, since the linear wave solution is not the exact solution to the fully nonlinear problem if taking into the non-Boussinesq terms in the Euler equation, the instability mechanism can also be initiated without a nonlinear free surface. 

We also would like to note that non-Boussinesq effects (and corresponding terms in the governing equation) can also lead to the generation of resonant super harmonics. In other words, harmonic generation can be obtained in uniform stratification with the rigid lid, if non-Boussinesq terms are not neglected. However, as will be shown shortly, the relative importance of such routing of energy is orders of magnitude less than that of nonlinear free surface avenue: Following the same procedure for getting \eqref{2011}, except not making Boussinesq approximation, we obtain,

\ba
&\rho \f{\p}{\p t}\lap w+\rho \lap_H(\vec{\b{u}}\cdot \g w)-\rho \f{\p^2}{\p x\p z} (\vec{\b{u}}\cdot \g u)-\rho \f{\p^2}{\p y \p z} (\vec{\b{u}}\cdot \g v)+g\lap_H\rho\nn\\
&+2\g\rho\cdot \g \f{\p w}{\p t}+ \f{\p w}{\p t}\lap\rho+2\g\rho\cdot \g (\vec{\b{u}}\cdot \g w)+\lap\rho\cdot(\vec{\b{u}}\cdot \g w)-\f{\p}{\p z}(\f{\p \vec{\b{u}}}{\p t}\cdot\g \rho)\nn\\
&-\f{\p}{\p z}[\g \rho\cdot (\vec{\b{u}}\cdot \g \vec{\b{u}})]-\f{\p}{\p z}(\f{\p \vec{\b{u}} }{\p t}\cdot\g \rho)=0.
\ea

Comparing the Boussinesq terms in the above equation and the quadratic terms on the free surface boundary condition, i.e., \eqref{402}, we find that their ratio is order of $a/k$. For the two cases presented above, the ratios are less than 1.6\%. \\

The harmonic generation through a nonlinear free surface involves two energy transfer processes simultaneously, in one process $(k,\omega)$ wave loses energy to $(2k,2\omega)$ wave and generates the second harmonic; in the other the $(2k,2\omega)$ wave forms a triad resonance with $(k,\omega)$ wave and sends energy back in an opposite way. The triad resonance condition is clearly satisfied since $2\omega-\omega=\omega$ and $2k-k=k$. If at the initial moment we only have $(k,\omega)$ wave (i.e. no second harmonic in the domain) then harmonic generation triumphs over triad resonance in transferring energy, resulting in the combined effect that energy only goes in one direction. But it can be shown that there are initial conditions combinations under which the energy at first goes from $(2k,2\omega)$ wave to the $(k,\omega)$ wave, but even in that case eventually the entire energy is transferred to the $(2k,2\omega)$ wave. This can be shown rigorously through Lyapunov stability theorem. Basically, the objective is to prove that $\A_1=0$ and $\B_2$ reaching its maximum is the asymptotically stable solution of the dynamical system described by equation  \eqref{200}. 

To prove this, we first note that \eqref{200} can be put in the following form
\ba
  \f{\d }{\d x} \lp \B_2^2 - \f{\alpha}{\beta}\A_1^2 \rp  = 0.
 \ea
Therefore, since $\f{\alpha}{\beta}$ can be shown to be always negative(see Appendix B), we have
\ba\label{3457}
  \B_2^2 - \f{\alpha}{\beta}\A_1^2 = \G^2,
\ea
where $\G$ is a constant that is determined by the initial conditions. Equation \eqref{3457} shows that $(\A_1,\B_2)$ are always moving on an ellipse. If we define a Lyapunov function $V(\A_1) = \G  - \alpha/|\alpha|[ \int_{0}^{x}\alpha \A_1(x)^2 dx+\B_2(0)]$, we find that $V(\A_1)\geqslant 0$ and $d V/d x<0$. According to  Lyapunov asymptotic stability theorem, the solution of the system converges to $\A_1=0$ from any starting points as $x$ goes to $\infty$.


\section{Conclusion}

Here we reported that certain internal gravity waves are inherently unstable and are not able to sustain their form. This new instability mechanism, a result of resonance harmonic-generation, draws the energy of an internal wave and hands it over to its second (or generally higher) harmonic. This resonance is distinguished from the classical triad resonance (and associated subharmonic instability) in that 1- a single wave may undergo the instability without requiring any external perturbation, and 2- the transfer of energy is irreversible and the original wave permanently loses its energy to its second harmonic. Extension of the results presented here to the third and higher harmonic generation is straightforward, but the strength of energy exchange in higher harmonics is much weaker. 

\appendix
\section{Derivation of the Interaction Equation}

Consider the propagation of waves in an inviscid, incompressible, adiabatic and stably stratified fluid of density $\rho(x,y,z,t)$, bounded by a free surface on the top and a rigid seafloor with a depth $h$ at the bottom. We consider a Cartesian coordinate system with $x,y$-axes on the mean free surface and $z$-axis positive upward. Equations governing the evolution of the velocity vector $\b{u} = \{u, v, w\}$, density $\rho$, pressure $p$ and surface elevation $\eta$ under Boussinesq approximation read
\bsa{901}
&\rho_0\f{D\b{u}}{Dt}=-\g p-\rho g \g z,~~-h<z<\eta\label{g101}\\
&\f{D\rho}{D t}=0,~~-h<z<\eta\label{g102}\\
&\g\cdot\b{u}=0, ~~-h<z<\eta\label{g103}\\
&\eta_{t}=\b{u}\cdot\g (z-\eta),~~z=\eta\label{g104}\\
&\f{D p}{D t}=0,~~z=\eta\label{g105}\\
&w=0,~~z=-h,\label{g106}
\esa
where $g$ is the gravitational acceleration, and $\rho_0=\bar{\rho}(z=0)$ is the mean density on the free surface with $\bar{\rho}(z)$ the background (unperturbed) density such that $\rho = \bar{\rho}(z)+ \rho' (x, y, z,t)$. Equation \eqref{g101} is the momentum equation (Euler's equations), \eqref{g102} comes from conservation of salt, and \eqref{g103} is continuity equation that together form five equations for five unknown variables of the problem (three components of velocity ${\bf u}$, pressure $p$ and density $ \rho$). Equations \eqref{g104}-\eqref{g106} are boundary conditions on the free surface and the bottom. 

Similar to the density perturbation $\rho'$, we define a pressure perturbation $p'$ via $p=\bar p(z)+p'(x,y,z,t)$ such that $d\bar p(z)/dz=-\bar{\rho}(z) g$. For the linear terms of the governing equation \eqref{g101} to be only in terms of $w$ we calculate
$\p/\p z[\g \cdot (\text{\ref{g101}})]-\lap (\text{\ref{g101}})_3$,
where (\ref{g101})$_3$  denotes the $z$ component of (\ref{g101})(likewise, 1,2 for $x,y$  will be used later). We obtain
\ba\label{2011}
\f{\p}{\p t}\lap w-\f{\p^2}{\p x\p z}\b{u}\cdot\g u-\f{\p^2}{\p y \p z }\b{u}\cdot\g v+\nabla_H^2(\b{u}\cdot\g w)+\f{g}{\rho_0}\nabla_H^2~ \rho'
=0,
\ea
where $\lap_{H}=\p^2/\p x^2+\p^2/\p y^2$ is the horizontal Laplacian. Taking the time derivative of \eqref{2011} and substituting $\rho'$ from the expansion of \eqref{g102}, i.e.,
\ba\label{g202}
\f{\p \rho'}{\p t}+\b{u}\cdot\g \rho'+w\f{d \bar{\rho}(z)}{d z}=0.
\ea
and denoting $N^2=-{g/\rho_0 \d \bar{\rho}(z)}/{ \d z}$ as the Brunt-V{\"a}is{\"a}l{\"a} frequency, we obtain 
%
%
\ba\label{401}
\f{\p^2}{\p t^2}\lap w+N^2 \lap_{H}w=
\f{\p^3}{\p x\p z\p t}\b{u}\cdot\g u+\f{\p^3}{\p y\p z\p t}\b{u}\cdot\g v-\nabla_H^2\f{\p}{\p t}(\b{u}\cdot\g w)+\f{g}{\rho_0} \lap_{H}(\b{u}\cdot\g\rho').
\ea

Expanding the surface dynamic boundary condition \eqref{g105} and keeping terms up to the second order we obtain
\ba\label{302}
\f{\p p'}{\p t}=w\rho_0 g-\b{u}\cdot\g p'-\f{\p^2 p'}{\p z \p t}\eta+\rho_0 g\eta \f{\p w}{\p z}+ g\eta w \f{\d \bar \rho}{\d z},~~~~~\rm{at}~~z=0.
\ea
where the last term can be equivalently written as $-N^2\bar\rho_0\eta w$.
Here $p'$ can be substituted from an expression obtained from taking the $x$ derivative of (\ref{g101})$_1$ added to  the $y$ derivative of (\ref{g101})$_2$:
\ba\label{301}
-\f{\p^3 w}{\p z\p t^2}+\f{1}{\rho_0}\lap_{H}\f{\p p'}{\p t}+\f{\p^2}{\p x \p t}\b{u}\cdot\g u+\f{\p^2}{\p y\p t}\b{u}\cdot\g v=0.
\ea
Substituting $\p p'/\p t$ from \eqref{301} in \eqref{302} we obtain
\ba\label{402}
\f{\p^3 w}{\p z \p t^2}-g\lap_{H}w=&\f{\p^2}{\p x \p t} \b{u}\cdot\g u+\f{\p^2}{\p y\p t}\b{u}\cdot\g v \nn \\
&+\lap_{H}\lp g\f{dw}{dz}\eta+N^2w\eta\rp-\f{1}{\rho_0}\lap_{H}\lp\b{u}\cdot\g p'+\f{\p^2 p'}{\p z \p t}\eta\rp.
\ea

Therefore the governing equation and boundary conditions correct to $\O(\ep^2)$ reduce to \eqref{401}, \eqref{402} and \eqref{g106}. For the ease of referring, we rewrite these equations here:
\bsa{300}
\f{\p^2}{\p t^2}\lap w+N^2 \lap_{H}w=&
\f{\p^3}{\p x\p z\p t}\b{u}\cdot\g u+\f{\p^3}{\p y\p z\p t}\b{u}\cdot\g v-\f{\p^3}{\p x^2\p t}\b{u}\cdot\g w \nn \\
&-\f{\p^3}{\p y^2\p t}\b{u}\cdot\g w+\f{g}{\rho_0} \lap_{H}\lp\b{u}\cdot\g\rho'\rp ~~~ -h<z<0,\label{3001}\\
\f{\p^3 w}{\p z \p t^2}-g\lap_{H}w=&\f{\p^2}{\p x \p t} \b{u}\cdot\g u+\f{\p^2}{\p y\p t}\b{u}\cdot\g v+\lap_{H}\lp g\f{dw}{dz}\eta+N^2 w \eta \rp \nn \\
& - \f{1}{\rho_0}\lap_{H}\lp\b{u}\cdot\g p'+\f{\p^2 p'}{\p z \p t}\eta\rp~~~~ z=0,\label{3002}
\\
w=&0,~~z=-h.\label{3003}
\esa
Equations \eqref{3001} and \eqref{3002} are identical to equations (A2) and (A7) of \cite{Thorpe1966}\footnote{except that we found three typos there: (1) for $S1$ in (A2), the sign of the last term with $g$ should be positive; (2) for $S3$ in (A7), the first term should be $\pmb{\omega}  \cdot \nabla w$; and (3) the term $g \nabla^2_1\frac{\partial}{\partial x}(\eta w_{z}) $ is missing. These are clearly typos as they do not appear in later expressions of \cite{Thorpe1966}.}.

To perform a weakly nonlinear analysis, we assume that internal waves in the system described above are small perturbations from the mean state of water at rest, i.e. $\bf u,\rho',p'\sim {\mathcal O}(\epsilon)$, where $\epsilon$ is a measure of the wave steepness. Considering the two-dimensional problem and assuming that the solution to this problem can be expressed in terms of a convergent series we define 
\ba
w(x,x_1,z,t)=\ep w_1(x,x_1,z,t)+\ep^2 w_2(x,x_1,z,t)+\O(\ep^3)
\ea
where $x_1=\epsilon x$ is the slow spatial variable and $\ep\ll1$ is a measure of steepness of the waves and $w_i\sim\O(1)$. Similar expressions exist for other variables, i.e. $u=\ep u_1+\ep^2u_2+\O(\ep^3)$, $\rho'=\ep \rho'_1+\ep^2\rho'_2+\O(\ep^3)$, $p'=\ep p'_1+\ep^2p'_2+\O(\ep^3)$ and $\eta=\ep \eta_1+\ep^2\eta_2+\O(\ep^3)$ with $u_i,\rho'_i,p'_i\sim\O(1)$ being functions of $x,x_1,z,t$, and $\eta_i\sim\O(1)$ being functions of $x,x_1$ and $t$.

Upon substitution into the governing equation, at the leading order $\O(\ep)$ we obtain
\bsa{g110}
&\f{\p^2}{\p t^2}\lap w_1+N^2 \lap_{H}w_1=0 &-h<z<0,\\
&\f{\p^3 w_1}{\p z \p t^2}-g\lap_{H}w_1=0& z=0,\\
&w_1=0, &z=-h.
\esa

At the second order $\O(\ep^2)$ we have

\bsa{120}
\f{\p^2}{\p t^2}\lap w_2+N^2 \lap_{H}w_2=&-2\lp\f{\p^2}{\p t^2}+N^2\rp \f{\p^2 }{\p x\p x_1}  w_1+\f{\p^3}{\p x\p z\p t}\b{u}_1\cdot\g u_1\nn \\
& +\f{\p^3}{\p y\p z\p t}\b{u}_1\cdot\g v_1-\f{\p^3}{\p x^2\p t}\b{u}_1\cdot\g w_1-\f{\p^3}{\p y^2\p t}\b{u}_1\cdot\g w_1\nn \\
&+\f{g}{\rho_0} \lap_{H}(\b{u}_1\cdot\g\rho'_1), \hspace{2cm} -h \leq z \leq 0.\label{121}\\
\f{\p^3 w_2}{\p z \p t^2}-g\lap_{H}w_2=&2 g \f{\p^2 w_1 }{\p x \p x1}+\f{\p^2}{\p x \p t} \b{u}_1\cdot\g u_1+\f{\p^2}{\p y\p t}\b{u}_1\cdot\g v_1\nn \\
&+\lap_{H}\lp g\f{dw_1}{dz}\eta_1+N^2 w_1 \eta_1\rp\nn \label{122} \\
&-\f{1}{\rho_0} \lap_{H}\lp\b{u}_1\cdot\g p'_1 +\f{\p^2 p'_1}{\p z \p t}\eta_1\rp, \hspace{1cm}z=0.\\
w_2=&0,\hspace{2cm} z=-h.
\esa

We now assume that waves with wavenumber and frequency $(k,\omega)$ and $(2k,2\omega)$ satisfy the internal waves dispersion relation (which is obtained from the linear equation \eqref{g110})
\ba
\D(k,\omega)=\omega^2-\f{gk}{\sqrt{ga/\omega^2-1}}\tan\lp kh\sqrt{ga/\omega^2-1} \rp,
\ea

 i.e. $\D(k,\omega)=0$ and $\D(2k,2\omega)=0$, and that they both exist in our domain of interest, though potentially with different amplitudes. The propagating wave solution to the linear equation \eqref{g110} then obtains as
\bsa{130}
w_1(x,x_1,z,t)=&A_1(x_1)\sin m_1(z+h)\sin{(kx-\omega t)}+B_1(x_1)\sin m_1(z+h)\cos{(kx-\omega t)}\nn \\
+&A_2(x_1)\sin m_2(z+h)\sin{(2kx-2\omega t)}+B_2(x_1)\sin m_2(z+h)\cos{(2kx-2\omega t)}
\esa
in which $A_1,A_2,B_1,B_2$ are amplitudes of each wave, $m_1^2=k^2(N^2-\omega^2)/\omega^2$ and $m_2^2=k^2(N^2-4\omega^2)/\omega^2$. Other variables $u,\rho',p'$ and $\eta$ can be found respectively via continuity equation \eqref{g103}, energy equation \eqref{g102}, kinematic surface boundary condition \eqref{g104} and the dynamic free surface boundary condition \eqref{g106}. 

The left-hand side of the second order equation \eqref{120} is identical in the form to the first order equation \eqref{g110}, but the right hand side of \eqref{120} is clearly non-zero and is a nonlinear function of the leading order solution $\u_1,\rho_1',p'_1,\eta_1$. It turns out, after substitution, that the right hand side contains terms with harmonics which are the same as the harmonics of the leading order equation (secular terms). A compatibility condition then must be enforced to make sure that the solution does not go unbounded, which is clearly unphysical. This compatibility condition determines the spatial behavior of the coefficients $A_i,B_i$. 

While the formulation presented here is general, our primary interest is when an initial wave with wavenumber and frequency $(k,\omega)$ \textit{resonates} its second harmonic $(2k,2\omega)$ whose initial amplitude is zero. Therefore in the following we use the adjectives \textit{original} and \textit{resonant} waves to refer to $(k,\omega)$ and $(2k,2\omega)$ waves respectively. We would like to emphasize that the formulation is general and works for any initial condition of the two waves. We will also comment that the presented approach can be easily extended for third and higher-harmonic generation.

Without loss of generality we assume that $B_1=0$, which only has to do with our choice of coordinate system. But we keep $A_2$ and $B_2$, since they determine the phase of the resonant wave $(2k,2\omega)$ with respect to the original wave $(k,\omega)$. The general solution to the second order problem takes the form
\bsa{140}
w_2(x,x_1,z,t)=&C_{11}(x_1,z)\sin{(kx-\omega t)}~~~+C_{12}(x_1,z)\cos{(kx-\omega t)}\\
+&C_{21}(x_1,z)\sin{(2kx-2\omega t)}+C_{22}(x_1,z)\cos{(2kx-2\omega t)}
\esa
where $C_{ij}(x_1,z)$'s are to be determined from \eqref{120}. Substituting \eqref{140} and \eqref{130} into \eqref{120}, and collecting same sine and cosine terms we obtain four ordinary differential equations for $C_{ij}$ ($i,j$=1,2):
\bsa{150}
-\om^2&C_{1i,zz}-m_1^2\om^2C_{1i}=E_{1i}, & -h<z<0,\\
-\om^2&C_{1i,z}+gk^2C_{1i}=F_{1i} & z=0,\\
&C_{1i}=0, & z=-h,
\esa \vspace{-0.5cm}
\bsa{160}
-4\om^2&C_{2i,zz}-4m_2^2\om^2C_{2i}=E_{2i}, & -h<z<0,\label{161}\\
-4\om^2&C_{2i,z}+4gk^2C_{2i}=F_{2i}, & z=0,\label{162}\\
&C_{2i}=0, & z=-h,\label{163}
\esa
where $E_{i1},F_{i1}$ are the coefficients of $\sin(ikx-i\om t)$, and $E_{i2},F_{i2}$ are the coefficients of $\cos(ikx-i\om t)$ in the right-hand side of \eqref{121} and \eqref{122} respectively. Let's first consider the equation for $C_{22}$, we obtain
\ba
E_{22}=-\f{4\om^2m_2^2}{k}\sin m_2(z+h) \f{\d A_2(x_1)}{\d x_1},~~~~ F_{22}=4gk\sin m_2h \f{\d A_2(x_1)}{\d x_1}
\ea
for which the solution to \eqref{161} that satisfies the boundary condition \eqref{163} is
\ba
C_{22}(x_1,z)=-\f{m_2}{2k}(z+h)\cos m_2(z+h) \f{\d A_2(x_1)}{\d x_1}.
\ea
Upon substitution into \eqref{162} we obtain
\ba
gk \lp  \f{2m_2h+\sin 2m_2h}{\cos m_2h} \rp\f{\d A_2(x_1)}{\d x_1}=0
\ea
therefore, since the coefficient is nonzero then $\d A_2/\d x_1\equiv$0. Physically speaking, this expression says that the amplitude $A_2$ does not change as waves propagate, or in other words, $A_2$ does not take part in the energy exchange. We, therefore, set $A_2$ equal to zero for the rest of the derivation.

We use the same approach as above for $C_{21}$. We have
\bsa{345}
&E_{21}=\f{4\om^2m_2^2}{k}\sin m_2(z+h) \f{\d B_2(x_1)}{\d x_1},\\
&F_{21}=-4gk\sin m_2h \f{\d B_2}{\d x_1}-2\lp 3\om m^2+\f{N^2k^2}{\om}\sin^2 m_1h\rp A_1^2(x_1)
\esa
for which
\ba
C_{21}(x_1,z)=\f{m_2}{2k}(z+h)\cos m_2(z+h) \f{\d B_2(x_1)}{\d x_1},
\ea
and upon substitution into \eqref{162} we obtain
\ba
\f{\d B_2(x_1)}{\d x_1}=\alpha A_1^2(x_1),
\ea
where
\ba
\alpha=-\f{6m_1^2\om\cos m_2h}{ g k(2m_2h+\sin 2 m_2 h) }.
\ea

For $C_{11}(x_1,z)$, we obtain $E_{11}=F_{11}=0$ and therefore the equation for $C_{11}(x_1,z)$ does not provide any extra information on $A_1(x_1)$ and $B_2(x_1)$. For $C_{12}$ we obtain
\ba
E_{21}=&\f{2m^2\om^2}{k}\sin m_1(z+h) \f{\d A_1(x_1)}{\d x_1}\nn\\
&+ \lcb I^{+}\sin\lb(m_1+m_2)(z+h)\rb+I^{-}\sin\lb(m_1-m_2)(z+h)\rb\rcb A_1(x_1) B_2(x_1) \\
F_{21}=&2gk\sin m_1(z+h) \f{\d A_1(x_1)}{\d x_1}+J A_1(x_1)B_2(x_1)
\ea
where
\bsa{180}
I^{+}&=\f{1}{8}\om(m_2+2m_1)(m_2^2+3k^2-m_1^2)\\
I^{-}&=\f{1}{8}\om(2m_1-m_2)(m_1^2-3k^2-m_2^2)\\
J&=\f{1}{4}\f{gk^2\lp 4m_1\cos^2 m_1h ~ \sin m_2 h-2m_2\cos m_2 h ~ \sin 2m_1 h  -3m_1\sin m_2h\rp}{\omega\cos m_1h}.
\esa
We obtain
\ba
C_{12}(x_1,z)=&-\f{m_1}{k}(z+h)\cos m_1(z+h) \f{\d A_1(x_1)}{\d x_1}\nn\\
&+\lcb \f{I^{+}\sin\lb(m_1+m_2)(z+h)\rb}{\om^2[m_1^2-(m_1+m_2)^2]}+\f{I^{-}\sin\lb(m_1-m_2)(z+h)\rb}{{\om^2[m_1^2-(m_1-m_2)^2]}}\rcb A_1(x_1) B_2(x_1).
\ea
Substituting into \eqref{162}, we obtain
\ba
\f{\d A_1(x_1)}{\d x_1}=\beta A_1(x_1)B_2(x_1),
\ea
where
\ba
\beta=-\f{k[\sin m_2 h(4m_1\cos^2 m_1h -3m_1) -2m_2\cos m_2h\sin 2m_1h]}{2\omega(2m_1h+\sin 2m_1h)}.
\ea

If we define the actual amplitudes $\A_1=\ep A_1$ and $\B_2=\ep B_2$ (note that $w=\ep w_1+\O(\ep^2)$)  then 
\bsa{900}
&\f{\d\B_2}{\d x}=\alpha \A_1^2\\
&\f{\d\A_1}{\d x}=\beta \A_1\B_2,
\esa
with the same $\alpha,\beta$ repeated here:
\bsa{1000}
\alpha&=-\f{6m_1^2\om\cos m_2h}{ g k(2m_2h+\sin 2 m_2 h) },\nn\\
\beta&=-\f{k[\sin m_2 h(4m_1\cos^2 m_1h -3m_1) -2m_2\cos m_2h\sin 2m_1h]}{2\omega(2m_1h+\sin 2m_1h)}.\nn
\esa

\section{Proof of the Sign of $\alpha/\beta$}

For the sign of $\alpha/\beta$, we have
\begin{equation}
\text{sign}(\frac{\alpha}{\beta})=\frac{\cos m_2 h}{\sin m_2 h (4 m_1 \cos^2 m_1 h-3 m_1)-2 m_2 \cos m_2 h \sin 2 m_1 h}.
\end{equation}

It is equvalent to the sign of
\begin{equation}
\tan m_2 h (4 m_1 \cos^2 m_1 h-3 m_1)-2 m_2  \sin 2 m_1 h.
\end{equation}
Rearranging terms of the above expression gives,
\begin{equation}\label{10001}
\cos^2 m_1 h (4 m_1\tan m_2 h -4 m_2\tan m_1 h )-3 m_1\tan m_2 h, 
\end{equation}

After substitution of the disperion relation,
\[
\omega^2=\frac{gk^2}{m_1}\tan m_1h
\]
and
\[
\omega^2=\frac{gk^2}{m_2}\tan m_2h,
\]

\eqref{10001} can be simplified to

\[
-3 m_1\tan m_2 h=-3 m_1 m_2\frac{\omega^2}{gk^2}
\]
which is less than zero always.

\newpage

\bibliographystyle{unsrt}

\end{document}